\documentclass[preprint,aps]{revtex4}
\input epsf
\begin{document}
\title{Modeling self-assembly of diblock copolymer-nanoparticle composites}
\author{F\'abio D. A. Aar\~ao Reis\footnote{On leave from Universidade Federal
Fluminense (Brazil). Email addresses: fdreis@wisc.edu, reis@if.uff.br.}
}
\affiliation{
Department of Chemistry, University of Wisconsin - Madison,
Wisconsin 53706, USA}
\date{\today}
\begin{abstract}
A cell dynamics method for domain separation of diblock copolymers (DBCPs)
interacting with nanoparticles (NPs) whose diffusion coefficients depend on
chain configuration is
proposed for self-assembly of DBCP/NP composites. Increasing NP
concentration slows down domain separation, but matching NP diffusion lengths
and lamellar size of DBCPs reduces this effect. The model also explains
features of different nanocomposites, such as morphological
transitions induced by NPs, the coexistence of lamellar and hexagonal patterns
in a single sample and peaked NP density profiles across the parallel domains.
\end{abstract}

\maketitle

\section{Introduction}

Mixtures of diblock copolymers (DBCPs) and nanoparticles (NPs) have attracted
much interest in recent years \cite{hamleyrev,bockstallerAM2005}. The
microphase separation of the DBCPs is expected to template NP arrangement,
which may improve the physical
properties of the composite or facilitate the production of other
nanostructures of technological interest.
However, the interaction of DBCPs and NPs may lead to a nontrivial
morphology even with low NP loadings
\cite{linNAT2005,grubbs,kimAM2005,balazsSCI2006,warren}, as anticipated by
models
of mixtures in thermodynamic equilibrium
\cite{thompsonSCI2001,leeMM2002}. This motivated intense theoretical
work \cite{zeng}, usually focusing on equilibrium properties. However, the most
highly ordered patterns are obtained in far from equilibrium conditions
\cite{lopesjaeger} and typical self-assemble times are of several hours or days.
Thus, understanding the pathways to form the desired structures is essential to
improve production methods.

This paper introduces a model for self-assembly of a mixture of DBCPs and
ex-situ formed NPs that describes features of various real composites. A cell
dynamics method (CDM) \cite{monica,hamleycds} represents the main
physico-chemical mechanisms of domain separation of DBCPs and provides a
realistic nanopattern morphology. The CDM variables, which represent the local
chain configurations, interact with NPs whose diffusion coefficients depend on
those configurations. Our CDM/NP model resembles those of Refs.
\protect\cite{balazsJPCB,ginzburg2000}, but it is a significant extension of
them because it accounts for a wider range of polymer-particle interactions.
Compared to other recent models, the advantage of
the CDM/NP approach is to address non-equilibrium and steady state features
simultaneously.

Among the non-equilibrium features represented by our model, emphasis is given
to the role of
particle mobility on the slow down of microphase separation and the effects of
NP concentration. High NP mobility is shown to destroy phase separation, while
very low mobility delays the alignment with an interacting surface by freezing
internal domain configurations. Connections with experimental work
\cite{jain,tangirala} are discussed and acceleration of global ordering is
suggested to occur when the NP diffusion lengths at the pure DBCP ordering
time matches the lamellar size.  The model also explains equilibrium features
observed in experiments, such as the peaked or broad distributions of NP
positions across DBCP lamellae \cite{chiu,kimMM2006}, the morphological
transitions induced by NP loading \cite{tangirala,sides,park,yeh}, and the
coexistence of different domain patterns in a single sample interacting with a
surface \cite{kimAM2005,compostoMM2007}.

The rest of this work is organized as follows. In Sec. II, we present the model
and discuss the simulation procedure. In Sec. III, we discuss the
non-equilibrium evolution of the DBCP/NP composites, with a focus on the case
in which the mixture interacts with an aligning surface. In Sec. IV, we show
how the increase of NP concentration may change the morphology of the
composites, in some cases producing samples where different structures
coexist. In Sec. V, we show the relation of NP diffusion coefficients and the
distribution of NPs across the ordered domains. In Sec. VI we summarize our
results and conclusions.

\section{Model and simulation procedure}

In the CDM, a field variable $\psi\left( \vec{r},t\right)$ at cell $\vec{r}$
represents the state of a nanoscopic region of a DBCP sample at time $t$. The
cell size is the length unit of the model and this is also the particle size, as
discussed below, thus it is expected to correspond to lengths between $2 nm$
and $10 nm$ in most applications (may be slightly larger in some cases). We will
work with lamellar sizes which are typically ten times larger than the cell
sizes, which in most cases correspond to high molecular weight polymers.

In a unit time interval, $\psi$ evolves as
\begin{equation}
\psi\left( \vec{r},t+1\right) = \langle\langle \psi\left( \vec{r},t\right)
\rangle\rangle + \Gamma\left( \vec{r},t\right) - \langle\langle
\Gamma\left( \vec{r},t\right) \rangle\rangle -
B\psi\left( \vec{r},t\right) + \xi\left( \vec{r},t\right) ,
\label{eqpsi}
\end{equation}
where $\Gamma$ contains short-range interaction terms,
$-B\psi\left( \vec{r},t\right)$ accounts for long-range interactions 
that restrict domain coarsening to a nanoscopic lengthscale, and $\xi$ is a
thermal (conservative) noise. $\tilde\Delta X\equiv \langle\langle X
\rangle\rangle-X$ is the discrete isotropized laplacian of $X$.
Here we follow the prescription of Ref. \protect\cite{ren} for simplifying the
short-range terms. In the case of symmetric DBCP, we choose 
\begin{equation}
\Gamma\left( \vec{r},t\right) =
f\left[ \psi\left( \vec{r},t\right)\right] + {\cal D}{\tilde\Delta
\psi\left( \vec{r},t\right)} - b {\tilde\Delta}^2 \psi\left( \vec{r},t\right) ,
\label{defgamma}
\end{equation}
with
\begin{equation}
f\left( \psi\right) = \left( 1+\tau\right)\psi -\tau\psi^3 .
\label{deff}
\end{equation}
The first contribution to $\Gamma$ represents attractive A-A and B-B
interactions that lead to segregation of blocks A and B of the DBCPs, since $f$
is a mapping with a flow to two symmetric stable points (it may be derived from
a $\psi^4$ approximation of a free-energy functional). The second term in Eq.
\ref{defgamma} is related to local diffusive dynamics and the third one
accounts for the curvature energy of the AB interface.  For a review on the
model and results, see Ref. \protect\cite{hamleycds}.

CDM is a simplification of the physico-chemical processes of coarsening systems:
the interaction terms have a simple form just to
push $\psi$ to the stable points, the same diffusion coefficient is
assumed for all chain configurations etc. Thus, it represents universal scaling
properties of DBCP domain separation, particularly those morphological
properties measured on lengthscales near and above the lamellar size,
but CDM is not expected to predict reasonable values of microscopic quantities
such as local chain densities.

Most works assume that $\psi$ is a difference of volume fractions of A and B
($\phi_A-\phi_B$) in a cell. However, this assumption makes typical CDM results
inconsistent with properties of strongly segregated DBCPs (SSDBCPs). For
instance, Fig. 1a shows the $\psi$ profile obtained from CDM using typical
parameters of the literature \protect\cite{monica,ren,hamleycds}. It is clear
that $|\psi|$ is well below $1$ in the middle of some domains (see
also Ref. \protect\cite{ren}). Instead, SSDBCPs
have large regions with $100\%$ A or $100\%$ B separated by narrow interfaces,
thus $|\phi_A-\phi_B|=1$ in the middle of those domains.
One idea to overcome this problem is to work with parameter ranges corresponding
to much lower temperatures, e. g. large values of $\tau$ (see  Refs.
\protect\cite{ohta,ren}). However, this produces frozen configurations with
very small domains (nearly the cell size) for extremely long times.

On the other hand, the pattern morphology of real SSDBCPs is well represented by
separating domains of positive and negative $\psi$ in CDM lattices shown in
the literature - e. g.
compare images in Refs.  \protect\cite{hamleyrev,darling} and CDM patterns in
Refs. \protect\cite{monica,ren,hamleycds}. This feature and the universality
character of CDM (i. e. the fact that CDM always includes the basic
physico-chemical mechanisms of the coarsening process of SSDBCPs)
lead us to propose a different interpretation of the field
variable: (i) the sign of $\psi$ represents the type of chain in a
cell, positive for A and negative for B; (ii) the absolute value of $\psi$ is
related to the local density at the cell and characterizes a certain position
across the domain ($\psi$ may be a non-monotonic function of the local density,
to be defined for a specific application); (iii) a range of small values of
$|\psi|$ (as defined below) characterizes interfacial and/or mixed regions.

The NPs are assumed to occupy approximately the whole volume
of a cell. The state $\psi$ of the CDM cell containing a NP represents
neighboring chains interacting with it, i. e. the NP environment. The NP
surface chemistry and this environment are responsible for the creation of
energy barriers for the NP to move \cite{lopesjaeger,gopinathan}. The NP
diffusion is assumed to take place in nearly equilibrium conditions with the
local environment: the higher the barrier, the lower the NP mobility, which
indicates a more stable configuration for NP localization. Thus,
a NP at position $\vec{r}$ executes random walks among neighboring cells with a
diffusion coefficient $D\left[ \psi\left( \vec{r},t\right)\right]$, i. e. the
rate with which the NP moves to neighboring cells is determined only by the
chain configuration at the cell it moved from (and, of course, by the NP
properties). Target cells for the NP steps are randomly chosen among the
nearest neighbors. Particle-particle interaction is restricted to the excluded
volume condition.

However, the presence of a NP in a cell also affects the states of neighboring
chains. It is expected that those states evolve not only according to the basic
(symmetric) CDM mechanisms, but also that neighboring chains move towards a
certain configuration $\psi_P$ that best wets the NP. For this reason, at each
time step the state of the NP cell is replaced by
\begin{equation}
\psi_{NP}\left( t+1\right) =\langle\langle
\psi \rangle\rangle + q\left( \psi_P-\langle\langle \psi \rangle\rangle\right) ,
\label{psinp}
\end{equation}
where the first term represents the current neighborhood of the
NP and the second one plus the CDM rules represent biased
chain diffusion towards the state $\psi_P$. In this context, $q$ measures the
bias intensity. Following this reasoning, the states of NN and next NN (NNN) 
cells change as
\begin{equation}
\psi_{NN} \to \psi_{NN}+q/6 \left( \psi_P-\psi_{NN}\right)
\label{psinn}
\end{equation}
and
\begin{equation}
\psi_{NNN} \to \psi_{NNN} +q/12 \left( \psi_P-\psi_{NNN}\right) ,
\label{psinnn}
\end{equation}
respectively (conservation is ensured by suitable changes in their
neighborhood).

Although the parameters $D$, $q$ and $\psi_P$ were interpreted
separately, they are related to the same microscopic NP-DBCP interaction, 
which is determined by DBCP properties and surface chemistry of NPs.
For a given application, those parameters cannot be viewed as independent
variables. However, the diversity of possible interactions in different systems
and the difficulty to predict the real interaction terms suggest to explore
them as independent for qualitative applications (in cases of solvent
annealing, those
parameters may also represent interactions with the solvent). For comparison we
recall that previous CDM/NP models assumed that NPs were permanently embedded in
a $\psi=1$ shell and had constant diffusion coefficient
\cite{balazsJPCB,ginzburg2000}, thus showing the extension
of the present approach.

Our simulations were done in square lattices, so that relatively large systems
could be simulated up to long times with a reasonable computational effort
(this is necessary due to the large number of parameters to control).
A disordered configuration at $t=0$ is represented by random distributions of
$\psi$ in the range $\left[ -0.025,0.025\right]$.
In some cases we used periodic boundaries, but a large fraction of our
simulations use a fixed boundary condition (FBC), where the row $y=0$ always has
$\psi=1$, in order to represent a material preferentially wet by block A.

A suitable set of CDM parameters (without NPs) were chosen to give the typical
$\psi$ profile shown in Fig. 1b after a domain separation time of order
$\tau_{SEP}\sim{10}^5$. This large value of $\tau_{SEP}$ is interesting because
it allows the study of the interplay between time scales of different
processes. 
In Fig. 1b, the cells closest to the A-B interface have $-0.1\leq \psi\leq 0.1$,
thus this interval is associated to interfacial and/or mixed regions. The
middle of A and B domains are characterized by $|\psi|\approx 0.3 - 0.4$ and
the lamellar width is $\lambda\approx 10$. The domain separation pattern
obtained under these conditions, shown in Fig. 1c, is similar to those
presented in the literature, typically obtained with the parameters of Fig. 1a.

Five simple models for $D\left( \psi\right)$ are used to investigate effects of
NP mobility.

Letting $D_A$, $D_B$, and $D_I$ stand for A, B, and
interfacial regions, respectively, the first three models are illustrated in
Fig. 2a and defined as: DM1,
with $D_B=\frac{1}{2}$, $D_I=\frac{1}{4}$, $D_A=\frac{1}{8}$;
DM2, with $D_B=\frac{1}{2}$, $D_I=\frac{1}{20}$, $D_A=\frac{1}{200}$;
DM3, with $D_B=\frac{1}{20}$, $D_I=\frac{1}{200}$, $D_A=\frac{1}{20000}$
(these values are given in time and length units of CDM defined above). DM1
represents large mobility in both domains, while DM2 and DM3 have
$D_B/D_A\sim {10}^2-{10}^3$ [as a rough guide, these are
the ratios for $Ag$ NPs in the domains of poly(styrene-b-methylmethacrylate)
(PS-PMMA) of Ref. \protect\cite{lopesjaeger}]. This form of
$D\left( \psi\right)$ is useful to test the effect of different particle
mobilities in different domains independently of possible intra-domain
variations of $D\left( \psi\right)$. With these models, any value $\psi_P\geq
0.1$ is reasonable for representing the effects of NPs on the chains because NP
mobility is uniformly low in the whole A domain, where $\psi\geq 0.1$.

Fig. 2b shows the other two diffusion models, where 
$D\left( \psi\right)$ is significantly reduced in certain chain
environments, simulating deep energy valleys for NP diffusion at those
positions. For consistently describing particle-polymer interaction, the
preference of the NP for a certain state [minimum of $D\left( \psi\right)$]
must be accompanied by a flux of the neighboring chains towards that state.
Thus, models DM4 and DM5 are respectively accompanied by $\psi_P=0.35$ (middle
of the A domains - see Fig. 1b) and $\psi_P=0.1$ (A side of the interfaces).

\section{Non-equilibrium self-assembly of the composite}

First we consider the effects of NP mobility on the global alignment
of the composites with the FBC.

In Fig. 3a, we show a FBC configuration without NPs obtained at $t={10}^6$,
which has $\psi$ profiles similar to Fig. 1b except very close to $y=0$.
In Figs. 3b-d we show FBC configurations at the same time
with $0.4\%$ of NPs and diffusion models DM1, DM2 and DM3, respectively.
The effect of NPs on the DBCPs is strong in this case, i. e. $q$ is large. It is
clearly observed that these small NP
concentrations can slow down domain separation or
eventually suppress the ordering.

The quality of the alignment with the $y=0$ border is
characterized by the correlation function along the $y$ direction,
\begin{equation}
C_y (y,t)\equiv \langle \psi\left(x',y',t\right)
\psi\left(x',y'+y,t\right)\rangle / \langle
{\left[\psi\left(x',y',t\right)\right]}^2 \rangle ,
\label{cy}
\end{equation}
with the average taken over $x'$, $y'$ and various realizations. The linear fits
of the maxima of $\log{|C_y (y,t)|}\times y$ plots give the corresponding
correlation lengths $L_y$. Fig 4a shows
the time evolution of $L_y$ in the same conditions of Fig. 3a-d.
Increasing $L_y$ indicates better alignment of the DBCP lamelles with the
interacting border, i. e. a faster global organization of the composite. From
Fig. 4a, it is clear that the best alignment with NPs is obtained with
diffusion model DM2, followed by DM3 and DM1.

In order to understand these features, an useful quantity is the diffusion
length of NPs inside block A at a certain time $t$, which is given by
\begin{equation}
l_D={\left( D_A t\right)}^{1/2} .
\label{ld}
\end{equation}
It measures the typical distance swept by a NP after that time if it is confined
to the A domain, where it has the lowest mobility.

For DM1, $L_y$ seems to saturate at long times (Fig. 4a), indicating that only a
finite region near the bottom surface is aligned, while the rest of the sample
is disordered. Simulation without the aligning surface provides a completely
disordered lattice, i. e. there is no phase separation even at long times.
$\psi_P=0.35$ and $q=0.1$ were used in those simulations, so that the A chains
are rapidly pulled to the NP neighborhood, but the fast movement of the NPs do
not allow the chains to reorganize around them. Indeed, at the separation time
$\tau_{SEP}$ of the pure system, we have $l_D\approx 112$, which is much
larger than the lamellar size $\lambda$ and explains the disordering effect of
NPs. These NPs are equivalent to a thermal noise that
brings the system above an order-disorder transition (ODT). 
It parallels the finding of Ref. \protect\cite{jain} that the ODT temperature
of poly(styrene-b-isoprene) (PS-PI) decreases as the concentration of mixed
silica particles increases.

On the other hand, very small mobility (DM3) does not disturb phase
separation in the middle of the lattice, as shown in Fig. 3d. Indeed, fixed NPs
may improve local organization, as discussed in Ref. \protect\cite{lee1999}.
However, they slow down the alignment with the border $y=0$ because they may
freeze a local order which does not match the order near the bottom surface.
The slowly increasing global ordering is confirmed by the slow increase in
$L_y$ of Fig. 4a. The NP diffusion length inside the A domain at $\tau_{SEP}$
is $l_D\approx 2$, while $l_D\approx 7$ at $t={10}^6$ (Fig. 2d). These values
of $l_D$ are both smaller than the lamellar width $\lambda\approx 10$,
justifying the interpretation of these slow particles as pinning centers of a
local domain configuration. The internal domain separation much faster than the
global ordering of the sample was experimentally observed in composites of
poly(styrene-b-2-vinylpyridine) (PS-P2VP) and $CdSe$ particles \cite{tangirala}.

The fastest global ordering (largest $L_y$) is obtained with DM2, where NPs in
block A have $l_D\approx 22$ at $\tau_{SEP}$. This value of $l_D$ is near the
domain period $2\lambda$, thus NPs can solidarily move with the neighboring
chains to the most favorable positions without generating undesired noise.
Other values of $q$ and $\psi_P$ lead to the same result. Thus, matching
NP diffusion length and lamellar period at the pure DBCP ordering time
$\tau_{SEP}$ may reduce the organization time $\tau_C$ of the composite with
the same morphology. In general, increasing temperature increases $D$ and
$\tau_{SEP}$, thus it also increases $l_D$ (but may decrease $\tau_C$ by
bringing it closer to $\tau_{SEP}$). This may be crucial for technological
applications where the experimental separation times upon annealing are
several hours or days.

As expected, the effect of increasing NP concentration is to increase the time
for global alignment, or reduce $L_y$ at a fixed time. This is shown in Fig. 4b
for diffusion model DM3 (very small NP mobility).
It parallels the increase of the  characteristic time for lamellar
alignment of PS-PMMA films with $Ag$ particles reported in Ref.
\protect\cite{compostoMM2007}.

In Figs. 4a and 4b, we note that $L_y$ scales diffusively with time (i. e. as
$t^{1/2}$) in the pure DBCP, but the presence of NPs seems to lead to slower
coarsening. It is not clear for us if there is a change in the universality
class of this non-equilibrium system due to the NPs or if this is a consequence
of corrections to the main scaling form. This may be an interesting point
for future theoretical investigation, but certainly much more accurate data in
larger system sizes would be necessary.

\section{Morphological changes}

The increase of NP concentration also has drastic effects on the composite
patterns at long times. The steady states of the CDM/NP model are able to
represent experimental features of some composites in thermodynamic
equilibrium.

First we analyze the effect of NP concentration with no aligning border.
Configurations at $t={10}^6$ are shown in Figs. 5a-c, respectively without NP
and with $0.8\%$ and $3.2\%$ of NPs. The pure DBCP has highly parallel lamellae
(Fig. 5a) and this structure has only small deformation with low NP
concentration (Fig. 5b). However, it changes to
a hexagonal structure with $3.2\%$ of NPs (Fig. 5c), even with weak effects of
the NPs on the chains ($q=0.02$). This is not a lattice effect because the CDM
used square grids. This is a morphological transition caused by the
stretching of the A domains that wet moving NPs, and can be achieved with
smaller concentrations for larger values of $q$. 

Such transitions were experimentally observed in Ref.
\protect\cite{sides} with PS-covered $Au$ particles in PS-P2VP.
In Ref. \protect\cite{park}, metallic and magnetic NPs changed the morphology of
PS-PI from cylindrical to spherical domains; the two-dimensional projections of
those patterns correspond to the lamellar-hexagonal transition illustrated in
Figs. 5a-c (see Fig.
2 of Ref. \protect\cite{park}). Refs. \protect\cite{sides} and
\protect\cite{yeh} show the opposite transition, from hexagonal to lamellar
morphology, but these are also cases where NPs stretched one of the domains.
Finally, the hexagonal morphology of PS-P2VP films was preserved with
high loadings of $CdSe$ particles in Ref. \protect\cite{zou}, which
suggests that suitable NP coatings were able to reduce NP effects on the
chains.

Even before these experiments were performed, NP-induced morphological
transitions were theoretically predicted using different thermodynamic
equilibrium models, e. g. in Refs.
\protect\cite{huh,thompsonSCI2001,leeMM2002,sides}.
The stretching of A domains by NPs was already shown in previous CDM/NP
models \cite{balazsJPCB,ginzburg2000}, but no morphological transition was
reported there. A recent work with dissipative particle dynamics (DPD)
\cite{heJPCB2008} only showed deformation of the lamellar structure of DBCP
with NPs. Thus, the novelty of the present approach is to show how those
transitions arise from a non-equilibrium cooperative dynamics.

The increase of NP loading may also lead to the coexistence of different
patterns in a single sample in contact with an aligning surface (FBC). This is
shown in Figs. 6a and 6b, where the samples have parallel lamellae near the
surface and nearly hexagonal order far from it. The increase of NP
concentration and the preferential location near the interfaces (i. e.
$\psi_P\approx 0.1$) favor the formation of the hexagonal structure, since the
NPs have larger diffusion lengths near the interfaces and more easily stretch
them. The concentration to obtain the coexisting patterns may vary, but a
minimum cooperative effort of the set of NPs (high $q$ or high concentration)
is required to produce it.
Similar structures were not shown in previous theoretical work, but it is
interesting to recall that Ref.
\protect\cite{pryamitsyn} showed evidence that non-selective NPs (which tend to
go to the interfaces) favor morphological transitions from a lamellar phase, as
illustrated here.

Different  patterns in a single sample were already observed experimentally.
Thick films of PS-P2VP containing PS-covered $Au$ particles
\cite{kimAM2005} showed a lamellar structure near the air-polymer interface and
a cylindrical structure (cross section with hexagonal packing) near the
substrate. The lamellar region propagated to longer distances compared to Figs.
6a-b, possibly affected by solvent evaporation and reduction of local NP
concentration - similar effects of solvent concentration were reported in Ref.
\protect\cite{compostoNL2007}. Moreover, in
Ref. \protect\cite{compostoMM2007}, different patterns in a single sample were
clearly observed: large $Ag$ particles froze lamellar configurations of
PS-PMMA perpendicular to the aligning (but relatively distant) substrate. The
particular NP location in that case was related to the interaction with the air
interface, but the overall effect of NPs with low mobility is similar to that
in Fig. 3d.

\section{NP distributions across aligned domains}

The NP position distributions across the parallel lamellae (FBC) can be also
related to NP diffusion.

If the NP diffusion coefficient is not very different in A and B domains (model
DM1), then large fractions of NPs are found in the B domains. For instance, for
low values of $q$ (e. g. $q=0.02$), a parallel lamellar organization is
obtained with more than $10\%$ of NPs in the B domains.

However, a significant localization of the NP inside the A domains is obtained
for models DM2 and DM3, where the mobility in the A domains is much smaller
than in the B domains.
In Fig. 7a, we show long time probability
densities $P\left( y\right)$ of finding a NP at position $y$ (measured
relatively to the center of a lamellae) for DM2 with different values of
$\psi_P$. In both cases the NPs are distributed almost
uniformly inside the A domains, with sharp concentration decrease at the
interfaces. Usually less than $1\%$ of the NPs are in the B domain and less than
$15\%$ in the interfacial region.

Similar shapes of $P\left( y\right)$ are found for a broad range of
parameters $q$ and $\psi_P$ and models DM2 and DM3. Thus, if the overall effect
of surface chemistry and polymer properties is to reduce NP mobility across the
whole A domain, then the NP distribution is broad, independently of the chain
configuration that tends to move to the NP neighborhood. In these cases, the
adjustments in chain configuration around the NPs (bias towards the state
$\psi_P$) are balanced out by thermal effects, while slow NP movement avoids
undesirable noise.

On the other hand, highly peaked $P\left( y\right)$ are obtained with DM4 and
DM5, as shown in  Fig. 7b. These are the cases where the NPs have severely
reduced mobility in a certain position of the A domains [minimum of
$D\left( \psi\right)$ - Fig. 2b]. This position determines the preferential
localization of NPs: the center of the A domains for DM4 and the interfaces for
DM5.

These results suggest that the localization of NPs strongly depends on the
surface chemistry being able to provide a favorable interaction with a
particular chain configuration, which may be the middle of the A domains or the
interface in the above examples. This favorable interaction corresponds to some
energy valley at that position, which reduces NP diffusion coefficient. Of
course additional conditions are important for the global organization, such as
reduced NP effects on the chains and avoidance of pinning effects due to very
low NP mobility, as discussed in Sec. III.

The above NP distributions resemble those of Kim et al
\cite{chiu,kimMM2006} for $Au$ particles in thick films of
PS-$b$-P2VP: preferential localization of the NPs in the center of
the PS lamellae or at the interface with P2VP, depending on the
density of coatings (PS or P2VP) of the NP surface. Recently, the same group
\cite{kimMM2008} showed NP position distributions similar to those in Figs. 7a-b
using various coatings. This suggests to search for an interpretation based on
position-dependent diffusion energy barriers during the annealing process.

\section{Summary and conclusion}

We proposed a cell dynamics model for DBCP domain separation
interacting with randomly moving NPs. Increasing NP concentration slows
down separation and global alignment, but this effect is reduced by matching
the lamellar size and NP diffusion lengths. The model also represents various
features of real composites, such as morphological
transitions induced by NPs, the coexistence of different patterns
in a single sample and various NP density profiles across the lamellae.

Previous works also predicted localized NP distributions
\cite{pryamitsyn,liu,jin,matsen,kang} and
morphological transitions \cite{thompsonSCI2001,leeMM2002,sides} in
thermodynamic equilibrium, but an advantage of the CDM/NP approach is to
explore the non-equilibrium routes leading to those and other properties.
An additional advantage is the low computational cost; for instance, one
configuration of a $256\times 256$ lattice at $t={10}^6$ is generated in less
than two hours in a desktop. This is important if one aims at testing hypothesis
theoretically before performing series of experiments.
Quantitative application is possible by associating
the time and length scales involved in the domain coarsening ( e. g.
$\tau_{SEP}$ and $\lambda$) to experimental values and choosing sensible
values of the other parameters. Certainly the model can also be extended, for
instance to include in-situ NP formation or NP interaction.

\vskip 0.5cm

{\par\noindent\bf Acknowledgment.} The author thanks Robert Hamers, Andrew
Mangham and Divya Goel for helpful
discussion, acknowledges support from CNPq for his visit to UW
and acknowledges support from CNPq and Faperj (Brazilian agencies) for his
simulation laboratory at UFF, Brazil.

\newpage

\begin{figure}[!h]
\leavevmode
\begin{center}
\epsfxsize = 15.0truecm
\epsffile{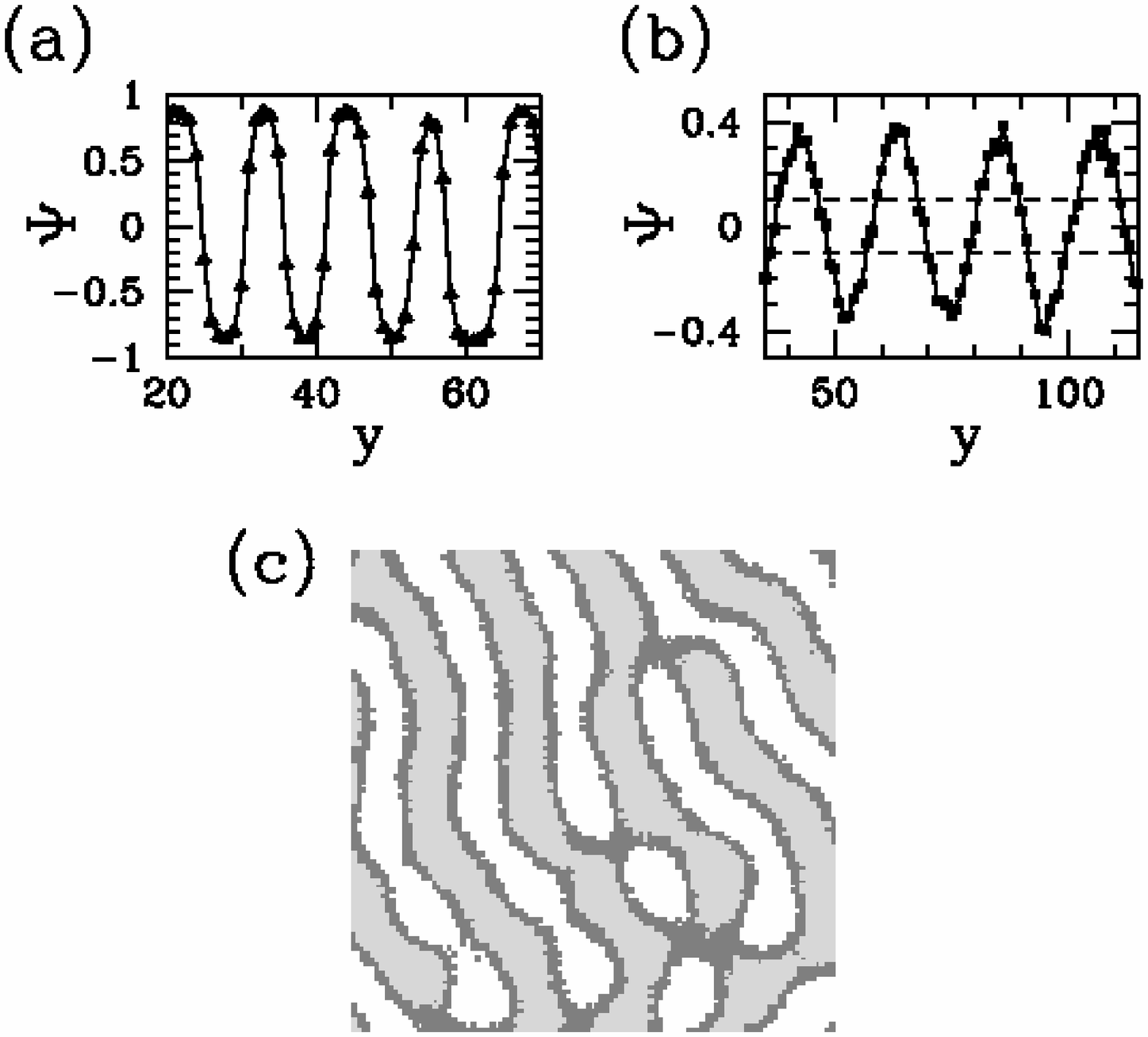}
\caption{(a) Long time $\psi$ profiles in directions perpendicular to the
parallel DBCP domains without NPs and traditional parameters of the literature:
$\tau =0.3$, $D=0.5$, $B=0.005$, $b=C=0$. (b) Profiles with typical parameters
of this work: $\tau=0.035$, $D=0.5$, $B=0.0005$, $b=0.35$, $C=0.02$; the region
between dashed lines is the interfacial and/or mixed region. (c) Typical lattice
configuration at $\tau_{SEP}={10}^5$ obtained with the CDM parameters of
this work. A (B) domains are light gray (white) and interfacial regions are 
medium gray.}
\end{center}
\end{figure}
\newpage

\begin{figure}[!h]
\leavevmode
\begin{center}
\epsfxsize = 15.0truecm
\epsffile{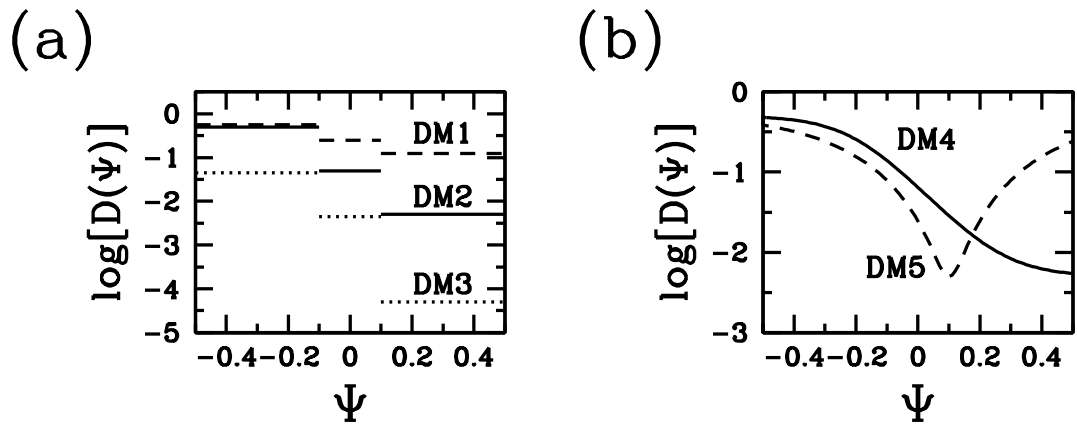}
\caption{Diffusion models: (a) DM1 (dashed lines), DM2 (solid lines), and DM3
(dotted lines); (b) DM4 (solid line) and DM5 (dashed line).}
\label{fig2}
\end{center}
\end{figure}
\newpage

\begin{figure}[!h]
\leavevmode
\begin{center}
\epsfxsize = 15.0truecm
\epsffile{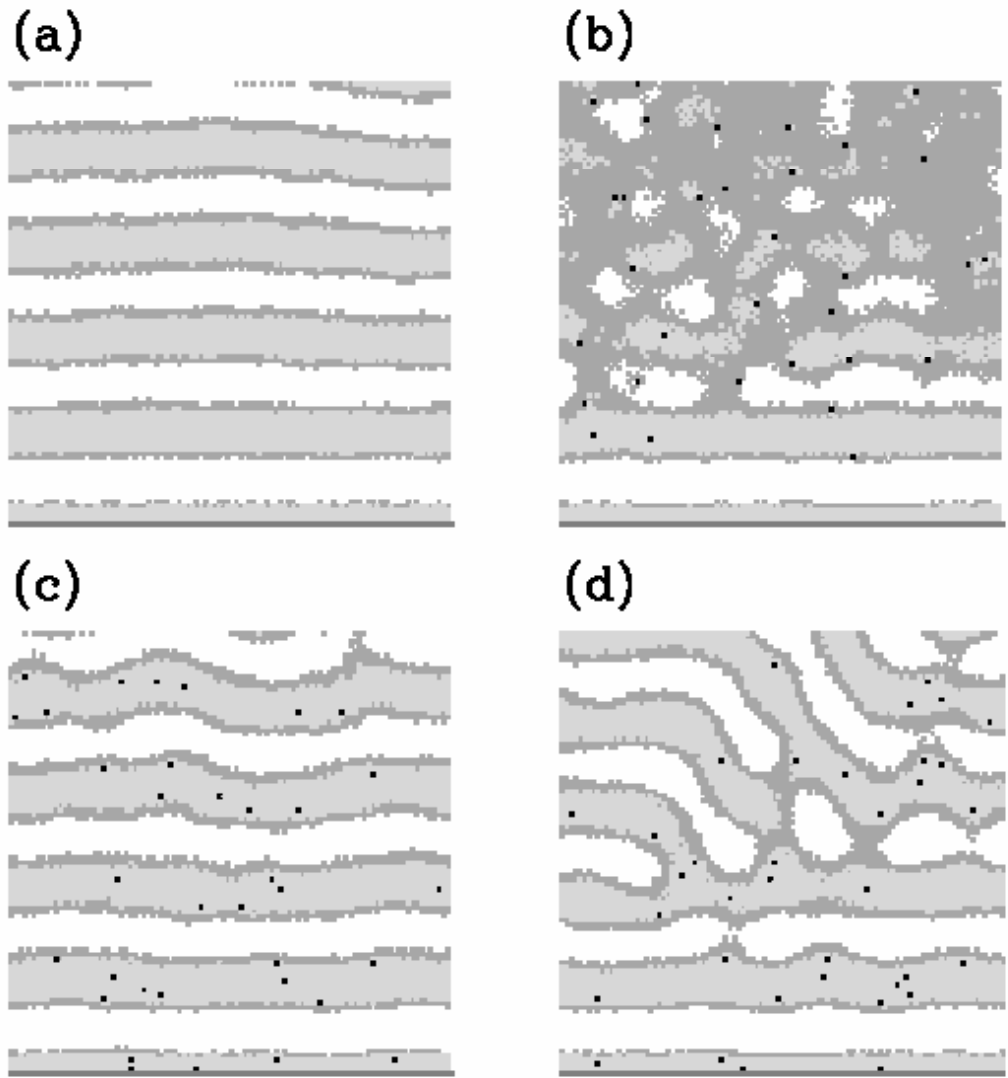}
\caption{Typical lattice configurations at $t={10}^6$ with the
FBC: a) without particles; b), c), d) $0.4\%$ of NPs with DM1,
DM2 and DM3, respectively, $q=0.1$,
and $\psi_P=0.3$. Surface with fixed $\psi=1$ is dark gray, A (B) domains
are light gray (white), interfacial regions are 
medium gray, and NPs are black.}
\label{fig3}
\end{center}
\end{figure}
\newpage

\begin{figure}[!h]
\leavevmode
\begin{center}
\epsfxsize = 15.0truecm
\epsffile{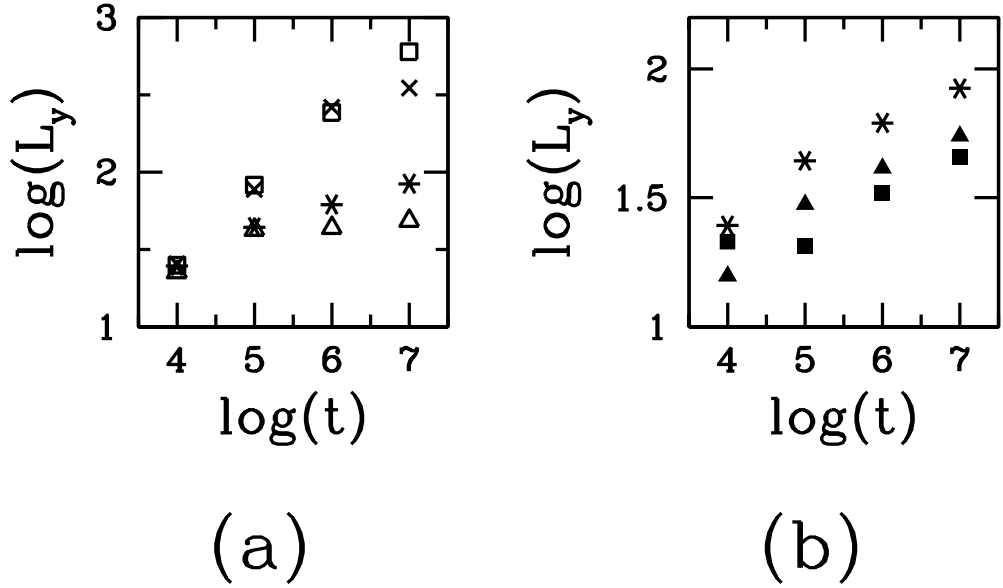}
\caption{Time evolution of correlation length with the FBC: (a) same
conditions of Fig. 2, with no particles (squares), DM1 (triangles),
DM2 (crosses), and DM3 (asterisks); (b) diffusion model DM3 with the same $q$
and $\psi_P$ of Fig. 2, and NP concentrations $0.4\%$
(asterisks), $0.8\%$ (filled triangles), and $1.6\%$ (filled squares). Error
bars are near the size of the points.}
\label{fig4}
\end{center}
\end{figure}
\newpage

\begin{figure}[!h]
\leavevmode
\begin{center}
\epsfxsize = 15.0truecm
\epsffile{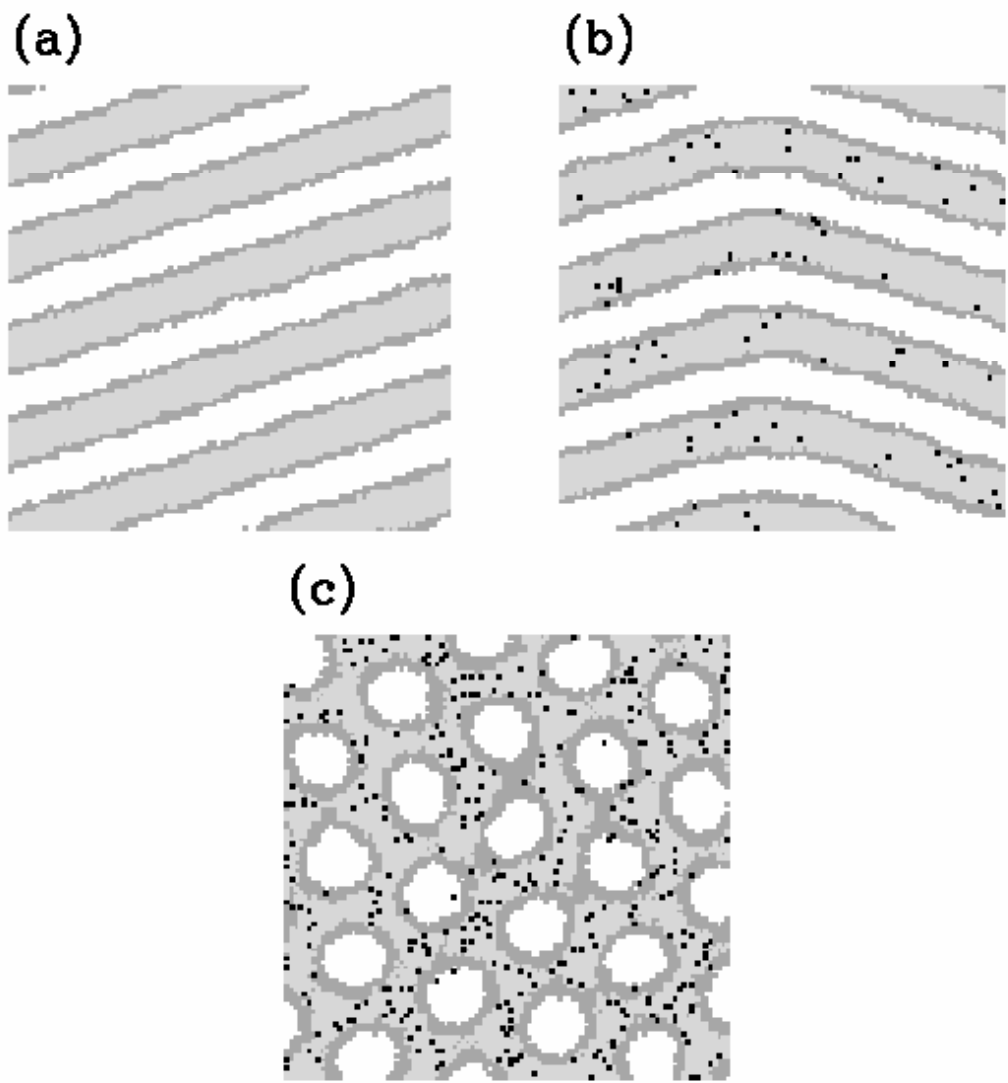}
\caption{Typical lattice configurations at $t={10}^6$ without (a), with $0.8\%$
(b) and with $3.2\%$ (c) of DM2 NPs, no aligning boundary, $b=0.05$, $q=0.02$,
and $\psi_P=0.3$. Color scheme is the same of Fig. 2.}
\label{fig5}
\end{center}
\end{figure}
\newpage

\begin{figure}[!h]
\leavevmode
\begin{center}
\epsfxsize = 15.0truecm
\epsffile{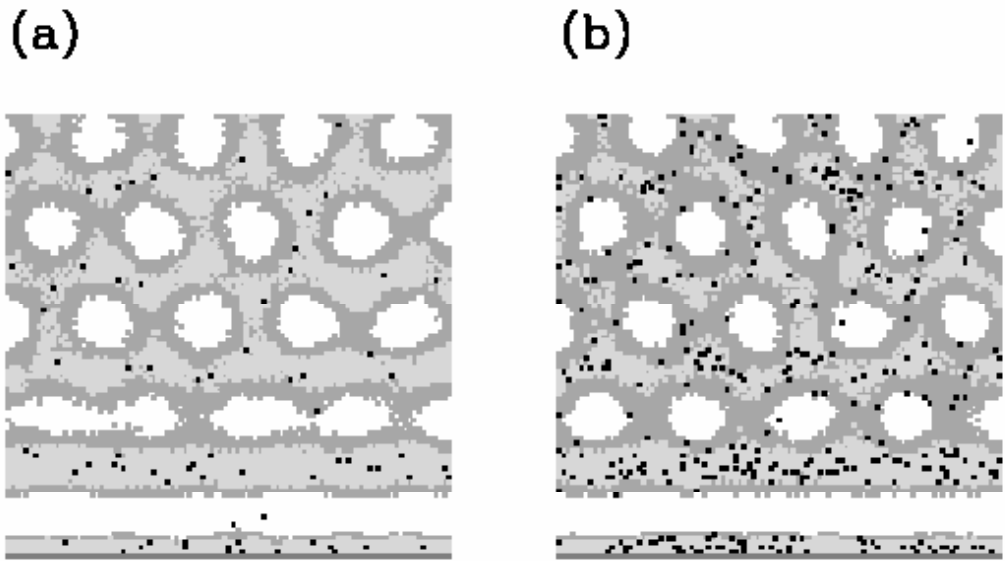}
\caption{Typical lattice configurations with FBC at $t={10}^7$ and: (a) $0.8\%$
of DM2 NPs, $b=0.35$, $q=0.1$, and $\psi_P=0.1$; (d) $3.2\%$ of DM5 NPs,
$b=0.35$, $q=0.02$, and $\psi_P=0.1$.
Color scheme is the same of Fig. 2.}
\label{fig6}
\end{center}
\end{figure}
\newpage

\begin{figure}[!h]
\leavevmode
\begin{center}
\epsfxsize = 15.0truecm
\epsffile{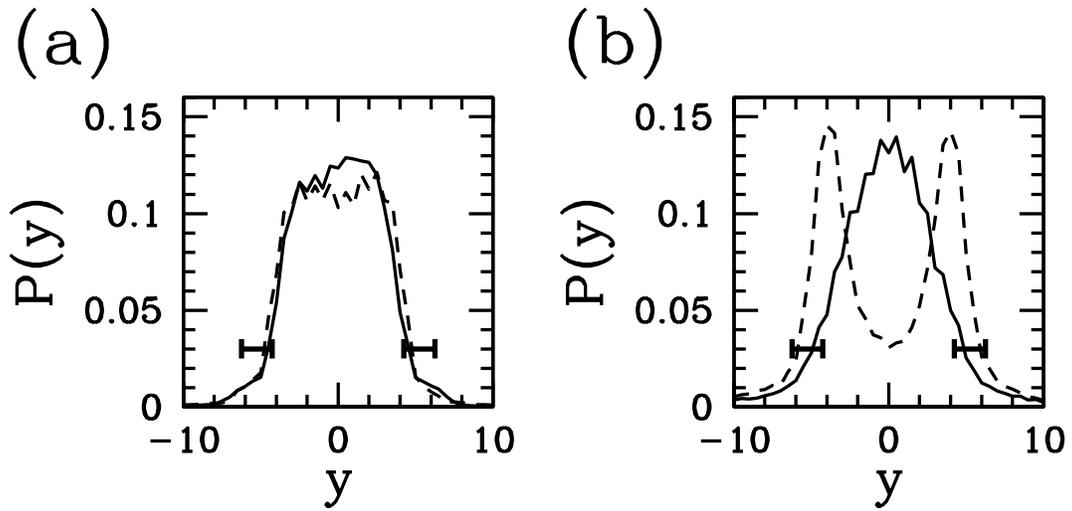}
\caption{NP position distributions across the parallel
lamellae for: (a) DM2 with $\psi_P=0.1$ (solid line) and $\psi_P=0.35$ (dashed
line); (b) DM4 with $\psi_P=0.35$ (solid line) and DM5 with $\psi_P=0.1$
(dashed line). Horizontal bars indicate typical interface regions.}
\label{fig7}
\end{center}
\end{figure}
\newpage

\end{document}